\documentclass[aps,prb,twocolumn,superscriptaddress]{revtex4-1}

\usepackage{epsfig}
\usepackage{amsmath}

\begin{document}


\title{Isotopic identification of engineered nitrogen-vacancy spin qubits in ultrapure diamond}


\author{T. Yamamoto}
\email[]{yamamoto.takashi@nims.go.jp}
\affiliation{National Institute for Materials Science, 1-1 Namiki, Tsukuba, Ibaraki 305-0044, Japan}
\affiliation{Japan Atomic Energy Agency, 1233 Watanuki, Takasaki, Gunma 370-1292, Japan}
\author{S. Onoda}
\affiliation{Japan Atomic Energy Agency, 1233 Watanuki, Takasaki, Gunma 370-1292, Japan}
\author{T. Ohshima}
\affiliation{Japan Atomic Energy Agency, 1233 Watanuki, Takasaki, Gunma 370-1292, Japan}
\author{T. Teraji}
\affiliation{National Institute for Materials Science, 1-1 Namiki, Tsukuba, Ibaraki 305-0044, Japan}
\author{K. Watanabe}
\affiliation{National Institute for Materials Science, 1-1 Namiki, Tsukuba, Ibaraki 305-0044, Japan}
\author{S. Koizumi}
\affiliation{National Institute for Materials Science, 1-1 Namiki, Tsukuba, Ibaraki 305-0044, Japan}
\author{T. Umeda}
\affiliation{Institute of Applied Physics, University of Tsukuba, 1-1-1 Tennodai, Tsukuba, Ibaraki 305-8573 Japan}
\author{L. P. McGuinness}
\affiliation{Institute for Quantum Optics, University of Ulm, D-89081, Ulm, Germany}
\author{C. M\"uller}
\affiliation{Institute for Quantum Optics, University of Ulm, D-89081, Ulm, Germany}
\author{B. Naydenov}
\affiliation{Institute for Quantum Optics, University of Ulm, D-89081, Ulm, Germany}
\author{F. Dolde}
\affiliation{3rd Physics Institute and Research Center SCoPE, University of Stuttgart, D-70174, Stuttgart, Germany}
\author{H. Fedder}
\affiliation{3rd Physics Institute and Research Center SCoPE, University of Stuttgart, D-70174, Stuttgart, Germany}
\author{J. Honert}
\affiliation{3rd Physics Institute and Research Center SCoPE, University of Stuttgart, D-70174, Stuttgart, Germany}
\author{M. L. Markham}
\affiliation{Element Six Limited, King's Ride Park, Ascot, Berkshire, SL5 8BP, United Kingdom}
\author{D. J. Twitchen}
\affiliation{Element Six Limited, King's Ride Park, Ascot, Berkshire, SL5 8BP, United Kingdom}
\author{J. Wrachtrup}
\affiliation{3rd Physics Institute and Research Center SCoPE, University of Stuttgart, D-70174, Stuttgart, Germany}
\author{F. Jelezko}
\affiliation{Institute for Quantum Optics, University of Ulm, D-89081, Ulm, Germany}
\author{J. Isoya}
\affiliation{Research Center for Knowledge Communities, University of Tsukuba, 1-2 Kasuga, Tsukuba, Ibaraki 305-8550, Japan
}


\date{\today}

\begin{abstract}
Nitrogen impurities help to stabilize the negatively-charged-state of NV$^-$ in diamond, whereas magnetic fluctuations from nitrogen spins lead to decoherence of NV$^-$ qubits. It is not known what donor concentration optimizes these conflicting requirements. Here we used 10\,MeV $^{15}$N$^{3+}$ ion implantation to create NV$^-$ in ultrapure diamond. Optically detected magnetic resonance of single centers revealed a high creation yield of $40\pm3$\% from $^{15}$N$^{3+}$ ions and an additional yield of $56\pm3$\% from $^{14}$N impurities. High-temperature anneal was used to reduce residual defects, and charge stable NV$^-$, even in a dilute $^{14}$N impurity concentration of 0.06\,ppb were created with long coherence times.
\end{abstract}


\maketitle

The realization of quantum registers, which are comprised of several quantum bits (qubits), is currently a central issue in quantum information and computation science.~\cite{Ladd} Among many competing quantum systems, photoactive defect spins of negatively charged nitrogen vacancy (NV$^-$) centers in diamond are unique solid-state qubits, due in part to ambient pressure and temperature operation.~\cite{Gruber,Jelezko1,Jelezko2} The NV$^-$ center is a single-photon emitter with zero-phonon-line (ZPL) at 637\,nm,~\cite{Davies} where both of $^3A_2$ electronic ground and $^3E$ excited states locate inside the diamond band-gap. The spin sublevels, $|{\it{m}}_{\rm{s}}=0\rangle$ and $|{\it{m}}_{\rm{s}}=\pm1\rangle$, of the  triplet ($S=1$) ground state are separated by $\sim2.87$\,GHz due to spin-spin interaction.~\cite{Reddy} Arbitrary states including superpositions of spin levels may be created by resonant microwave pulses after optical initialization, and then readout by measuring fluorescence intensity.~\cite{Jelezko1} Experimental proofs of strongly-coupled NV$^-$ spins,~\cite{Neumann,Dolde,Yamamoto1} magnetic coupling between a NV$^-$ spin and another electron spin~\cite{Gaebel,Yamamoto1} or nuclear spins,~\cite{Jelezko3,Dutt,Neumann2,Jiang,Robledo1}, in addition to coupling to photons~\cite{Togan,Buckley} or optical cavities,~\cite{Park,Englund,Faraon} exemplify the robust yet mutable nature of the NV scheme as well as the beginnings of scalability.

The NV quantum coherence decays in time due to magnetic fluctuations from substitutional nitrogen ($\text{N}_s^0$) electron spins and $^{13}$C nuclear spins, and spin-lattice relaxation.~\cite{Hanson1,Mizuochi,Markham} Thus, the use of high purity ($[\text{N}_s^0]\sim\text{ppb}$) type IIa diamonds with reduced $^{13}$C content, and position controlled N ion implantation to create NV$^-$ centers, is a promising avenue towards a high quality multi-qubit system.~\cite{Dolde} Nevertheless, substitutional nitrogen impurities, which donate electrons to NV centers, are actually essential for stabilizing the NV$^-$ charge state.~\cite{Mita} The negative NV charge state is predominant at thermal equilibrium if $[\text{N}_s]=[\text{N}_s^0+\text{N}_s^+]>[\text{NV}]$, and this is generally true for isolated NV centers in type Ib diamond $([\text{N}_s^0]\sim20-200$\,ppm).~\cite{Kennedy1,Kennedy2,Mason} With decreasing N$_s^0$ donor concentration, microscopic distributions of donors surrounding each NV center are significant for the charge state, rather than the Fermi position relative to the ground state of NV$^-$. As a result locally inhomogenous distributions of either NV$^0$ or NV$^-$ are expected.~\cite{Collins} This may explain the relatively large reduction in NV$^-$ population observed by photoluminescence spectroscopy in type IIa diamonds ($[\text{N}_s^0]\sim 30-300$\,ppb).~\cite{Kennedy2} The presence of the neutral NV$^0$ charge state ($S=1/2$) is undesirable as its applications are hindered by rapid dephasing in the ground state. Therefore, the understanding of a minimum concentration threshold of $\text{N}_s^0$ impurities in order to form stable NV$^-$ spin qubits is of concern for reliable engineering and scalability.

In this study we isotopically distinguish engineered $^{15}$NV$^-$ spin qubits due to $^{15}$N implantation from $^{14}$NV$^-$ due to preexisting $^{14}$N impurities in ultrapure diamond, both of which can be created by $^{15}$N$^{3+}$ (10\,MeV) implantation. Using a combination of confocal microscopy and spin resonance, we observe an implantation creation yield of $\sim100\%$, of which $^{14}$NV$^-$ centers comprise more than half this value. The nitrogen concentration of less than 0.1\,ppb is low enough to attain long coherence times ($\sim2$\,ms) and sufficient to stabilize the charge state of NV$^-$ qubits, under reduced concentrations of residual defects by high-temperature anneal. 

\begin{figure}
\includegraphics[width=8.6cm]{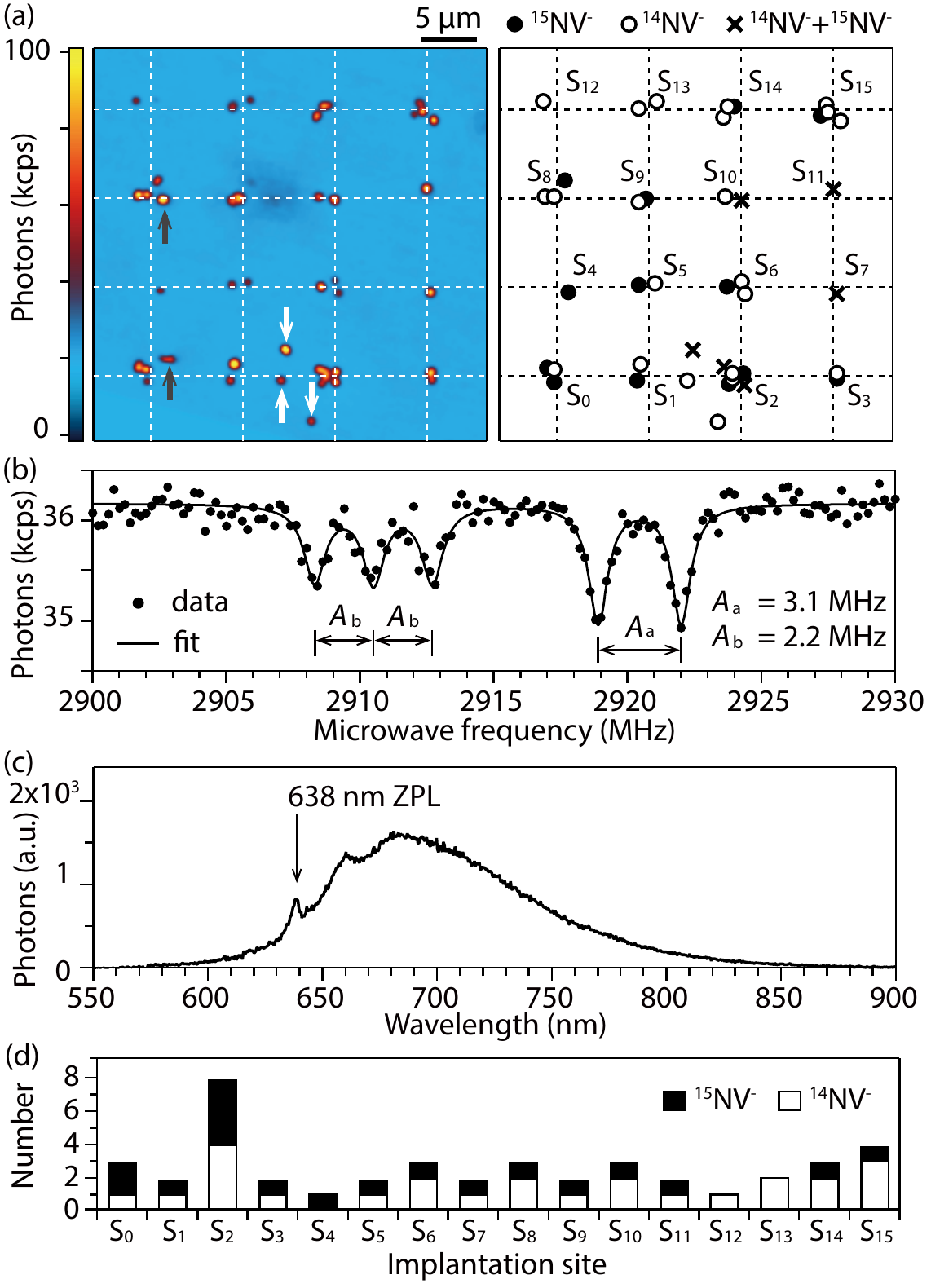}%
\caption{(Color online) (a) Confocal microscope image of lateral ($xy$ plane) distribution of NV$^-$ centers with 532\,nm excitation at a depth of $\sim3.8\,\mu$m (left). The white dashed lines indicate a calculated square grid of implantation sites (2.8\,ions/site).~\cite{TYS} The fluorescent spots indicated by black arrows were unknown centers (see text).
The map of NV$^-$ centers identified by ODMR (right): $^{15}$NV$^-$ (solid circle), $^{14}$NV$^-$ (open circle), and $^{14}$NV$^-$-$^{15}$NV$^-$ pairs (cross). (b) ODMR spectrum of a NV$^-$ pair comprised of $^{14}$NV$^-$ (hyperfine splitting of $A=2.2$\,MHz) and $^{15}$NV$^-$ ($A=3.1$\,MHz) in the transition of $|{\it{m}}_{\it{s}}=0\rangle\Leftrightarrow|{\it{m}}_{\it{s}}=+1\rangle$ (c) PL spectrum of a $^{15}$NV$^-$ center with 532\,nm excitation. (d) The number of $^{15}$NV$^-$ (black bar) and $^{14}$NV$^-$ (white bar) in each implantation site. \label{FIG1}}
\end{figure}

In experiments, a high-purity, 99.99\% $^{12}$C-enriched (0.01\%-$^{13}$C) homoepitaxial diamond (Element Six Ltd.) grown by chemical vapor deposition was used. The concentration of N impurities was expected to be less than 0.1\,ppb from the crystal growth condition,~\cite{Edmonds,Balasubramanian} which is far below the detection limit of secondary ion mass spectroscopy or electron spin resonance for the film thickness here. $^{15}$N$^{3+}$ ions with an incident energy of 10\,MeV per ion were implanted into the (100) crystal surface. By scanning a microbeam of full-width at half-maximum (FWHM) size $\sim1.5\,\mu$m, a square grid of implantation sites separated by $\sim8\,\mu$m was created [Fig.\,\ref{FIG1}(a)]. The average number of implanted ions was 2.8 per implantation site by measuring the beam flux before and after implantation.~\cite{Sakai,Kamiya} To form NV centers, the sample was annealed at $1000^{\circ}\textrm{C}$ for 2\,h in a vacuum of $\sim10^{-6}$\,Torr. 

Observation of hyperfine structure of either $^{15}$N (with a nuclear spin of $I = 1/2$, natural abundance 0.37\%) or $^{14}$N ($I=1$, 99.63\%) by optically detected magnetic resonance (ODMR) spectroscopy allowed determination of whether the investigated NV$^-$ centers were due to $^{15}$N implants or $^{14}$N impurities already present in the epitaxial layer [Fig.\,\ref{FIG1}(a)].~\cite{Rabeau} The ODMR spectra were measured at room temperature, under a static magnetic field of $\sim2$\,mT in order to separate the two transitions: $|{\it{m}}_{\it{s}}=0\rangle{\Leftrightarrow}|{\it{m}}_{\it{s}}=+1\rangle$ and $|{\it{m}}_{\it{s}}=0\rangle{\Leftrightarrow}|{\it{m}}_{\it{s}}=-1\rangle$. 
ODMR was able to resolve NV$^-$ pairs with different axis orientations when their separation is below the confocal resolution [Fig.\,\ref{FIG1}(b)]: the triplet and doublet hyperfine structures show $^{14}$NV$^-$ with hyperfine constant $A=2.2$\,MHz and $^{15}$NV$^-$ with $A=3.1$\,MHz, respectively.
No pairs consisted of two $^{15}$NV$^-$ centers, rather, all pairs were formed by one each of $^{15}$N and $^{14}$N. This agrees with a numerically estimated formation probability of $\sim1$\% for $^{15}$NV$^-$ pairs locating within a $\sim0.3~\mu$m laser spot, when given our microbeam size (FWHM of $1.5\,\mu$m).

Room-temperature photoluminescence (PL) spectra were also measured for individual centers with 0.5\,mW of 532\,nm excitation (into the objective) and an accumulation time of 30\,sec. Recent studies have shown that dynamical charge conversion between NV$^-$ and NV$^0$ occurs under illumination, and the controllable dynamics have been discussed.~\cite{Rittweger,Waldherr,Beha,Han,Siyushev,Aslam} Both NV$^-$ and NV$^0$ may be observed even for a single NV center, with time-averaged PL spectroscopy, if significant photoconversion appears during optical pumping (for example, see Ref.~\onlinecite{Gaebel2}). However, we observed characteristic spectrum of the negative charge state from all $^{14}$NV$^-$ and $^{15}$NV$^-$ centers:~\cite{TYS} a weak zero-phonon-line (ZPL) at 638\,nm accompanied with broad vibronic sidebands [Fig.\,\ref{FIG1}(c)]. No distinct signals of NV$^0$ charge states (575\,nm ZPL) were found, at least within the accumulation time of 30\,sec. Two centers that didn't show an ODMR signal, indicated by black arrows in Fig.\,\ref{FIG1}(a), were unknown centers since their PL spectra were different from either NV$^-$ or NV$^0$.

Additional spins belonging to paramagnetic residual defects, resulting from the implantation and anneal process, may dominate the decoherence of implanted NV$^-$ spins.~\cite{Yamamoto2,Naydenov2} Also, residual point defects such as divacancies may act as accepters,~\cite{Deak} to ionize NV$^-$ to NV$^0$. To overcome these obstacles, high temperature anneal has been shown to be effective in reducing the concentration of residual paramagnetic defects at $\ge1000^{\circ}\textrm{C}$,~\cite{Yamamoto2,Iakoubovskii3,Lomer} with a concomitant increased population of NVs with long coherence times\cite{Naydenov2} and improved spectral stability\cite{Chu} when compared to 800$^{\circ}\textrm{C}$ anneal. Previous studies have also investigated NV charge instability due to residual defects after ion implantation,~\cite{Meijer,Waldermann,Gaebel,Gaebel2,Rabeau,Naydenov1,Pezzagna1,Toyli,Schwartz} neutron~\cite{Mita} or electron irradiation,~\cite{Mason} and anneal temperatures of 600-900$^{\circ}\textrm{C}$, while effects from the surface are additionally involved in shallow implantation studies.~\cite{Santori,Rondin,Fu,Hauf,Ofori-Okai} In the present work, both high temperature anneal and high energy implantation were performed to provide a clean environment with minimal degradation of the NV$^{-}$ properties.

Figure~\ref{FIG1}(d) shows the number of $^{14}$NV$^-$ and $^{15}$NV$^-$ centers at each implantation site in Fig.\,\ref{FIG1}(a), labeled by S$_{\textit{j}}$ ($j=0, 1, 2, \cdots, 15$). In total, eighteen $^{15}$NV$^-$ and 25 $^{14}$NV$^-$ were created from 45 implanted $^{15}$N ions (2.8$\times$16) in the sixteen implantation sites. Dividing the number of created NV$^-$ centers by the number of implanted $^{15}$N ions gives a creation yield of $Y=40\pm3$\% for $^{15}$NV$^-$ and $56\pm3$\% for $^{14}$NV$^-$, where the NV$^-$ centers indicated by the white arrows in Fig.~\ref{FIG1}(a) were excluded for the counting since these centers are far from an implantation site and might be owing to a miss-hit. As reported in Ref.~\onlinecite{Yamamoto2}, the spin coherence times ($T_2$) in this implantation area were $\sim2$\,ms at room temperature, which are the longest among implanted NV$^-$ qubits, and comparable to the longest recorded for naturally-formed NV$^-$ centers during crystal growth.~\cite{Balasubramanian, Ishikawa, Jahnke}
In addition to a total yield of 96\%, we also obtained $\sim100\%$ yield for implanted NV$^-$ centers with $T_2$ times up to 1.6\,ms in another high purity $^{12}$C-99.99\% enriched diamond (Element Six Ltd.) by similar implantation and annealing process (data not shown).

Low-energy (10-30\,keV) nitrogen ion implantation has provided creation yields of 20-21\%,~\cite{Naydenov1,Yamamoto1} which is just below the maximum expected value of 25\%.~\cite{Antonov} Compared to this, high-energy (18\,MeV) implantation has exhibited 45\% yield, which was interpreted due to the increased number of vacancies generated by increasing the implantation energy.~\cite{Pezzagna1} However, NV centers comprised of preexisting N impurities may also be counted in the yield, and thus the creation efficiency should change with the concentration of $[\text{N}_s^0]$ in each sample.

To investigate further the creation of $^{14}$NV$^-$ and $^{15}$NV$^-$ centers, we measured the spatial distributions of each center by confocal microscopy. A diffraction-limited fluorescence spot from a single center has a lateral diameter ({\it{xy}}-plane) of $\sim0.3\,\mu$m and a diameter of $\sim0.7\,\mu$m along the optical axis ({\it{z}-}axis). To measure the coordinates, $(x, y, z)$, of individual NV$^-$ centers with high precision, we used a Gaussian fit to find the position of maximal intensity of the fluorescence profile, giving an accuracy of $<0.1\,\mu$m in all axis-directions. NV$^-$ pairs were inseparable by fitting and thus measured as at the same position.~\cite{TYS} We compared the observed spatial distribution of NV$^-$ centers, to the computed statistical distributions of implanted $^{15}$N atoms and vacancies using stopping and range of ions in matter (SRIM) Monte Carlo\cite{Ziegler} (a displacement energy of 37.5\,eV,~\cite{Koike} a diamond density of 3.52\,g/cm$^3$, and $8 \times 10^4$ of incident $^{15}$N ions were used). By using the vacancy distribution computed by SRIM, we then simulated a statistical vacancy distribution after diffusion with an isotropic diffusion length of $\sqrt{2Dt}\approx 0.08\,\mu$m, where $D=D_0\exp[-E_{\text{a}}/(k_{\text{B}}T)]$, with diffusion coefficient $D_0=3.7\times10^{-6}$\,cm$^{2}$/s (Ref.~\onlinecite{Hu}), Boltzmann's constant $k_{\text{B}}=1.4\times10^{-23}$\,T/K, activation energy $E_{\text{a}}=2.3$\,eV (Ref.~\onlinecite{Davies2}), temperature $T=1273$\,K, and time $t=7200$\,s.~\cite{TYS}

\begin{figure}
\includegraphics[width=8.6cm]{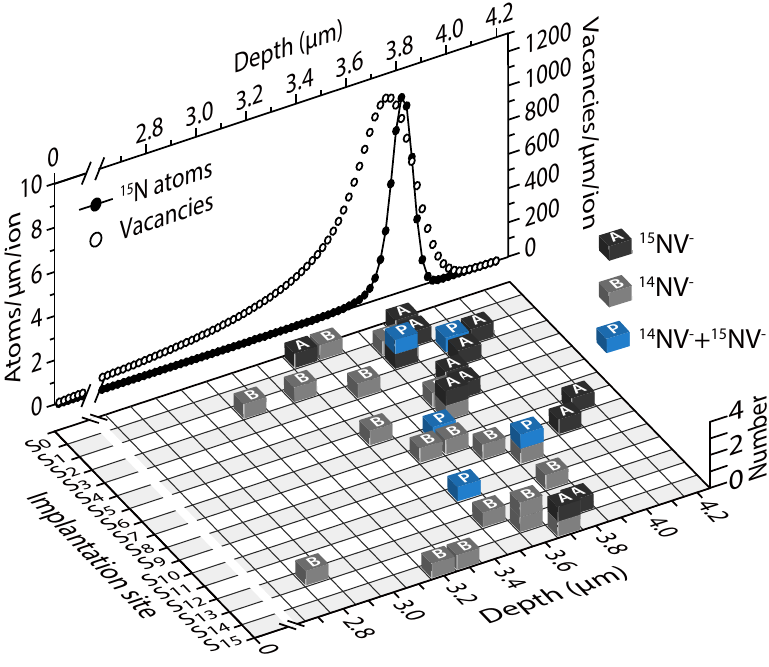}%
\caption{(Color online) (b) Depth ({\it{z}-}axis) distributions of observed NV$^-$ centers at each implantation site as histogram: $^{15}$NV$^-$ (black), $^{14}$NV$^-$ (gray), and $^{14}$NV$^-$-$^{15}$NV$^-$ pairs (blue). The graph in the back side shows the simulated depth distributions for implanted $^{15}$N atoms (solid circle) and vacancies with isotropic diffusion length of $\sim 0.08\,\mu$m (open circle). \label{FIG2}}
\end{figure}

Experiments showed no obvious difference in the lateral ({\it{xy}}-plane) distributions between $^{15}$NV$^-$ and $^{14}$NV$^-$ centers.~\cite{TYS} On the other hand, we observed differences in depth ({\it{z}-}axis) distributions between the two nitrogen sources: the mean depth of  $3.5\pm0.3\,\mu$m for $^{14}$NV$^-$ was shallower than that of $3.8\pm0.2\,\mu$m for $^{15}$NV$^-$. The depth distribution of $^{15}$NV$^-$ centers shows no evidence of channeling,~\cite{TYS} and the mean depth agrees well with the computed depth range of implanted $^{15}$N atoms by SRIM ($3.82\pm0.04\,\mu$m). Interestingly, all $^{14}$NV$^-$ centers were located at the same or shallower depths than those of $^{15}$NV$^-$ centers (Fig.~\ref{FIG2}), which is expected to result from the vacancy profile which trails towards surface (open circle in Fig.\,\ref{FIG2}), as compared to the sharp Bragg peak of implanted $^{15}$N atoms (solid circle). No $^{14}$NV$^-$ centers were observed at depths shallower than $2.8\,\mu$m due to the low concentration of vacancies. 

Now we discuss the creation efficiency for implanted $^{15}$NV$^-$ centers. The creation yield, $Y(c_v)$, is written as
\begin{equation}
Y(c_v)=\frac{N_q(c_v)}{N_i}=\frac{{N}_i\times P_r}{N_i}\times P_t(c_v),
~\label{F1}
\end{equation}
where $N_q/N_i$ is the ratio of the number of implanted $^{15}$N ions ($N_i$) to the number of $^{15}$NV$^-$ created by implantation ($N_q$). The product ${N}_i\times P_r$ gives the number of substitutional $^{15}$N atoms among the implanted $^{15}$N ions, dependent on the replacement probability $P_r$ at a carbon lattice site, and $P_t(c_v)$ is the probability of trapping a vacancy at a neighboring site of substitutional $^{15}$N atom, which is proportional to concentrations of vacancies, $c_v$.~\cite{Pezzagna1} For $P_t(c_v)\approx 1$, the obtained yield of 40\% gives a replacement probability of $P_r\approx 0.4$, which agrees well with recent molecular dynamics simulations showing that 37\% of implanted ions are substituted to lattice sites after implantation.~\cite{Antonov} This suggests that our 10\,MeV implantation and annealing process provides enough vacancies to form $^{15}$NV$^-$ with a high trapping probability of $P_t(c_v)\approx 1$, and the creation efficiency is limited by those $^{15}$N implants which remain as interstitial nitrogen after implantation ($\sim60\%$ of implanted ions). 

\begin{figure}
\includegraphics[width=8.6cm]{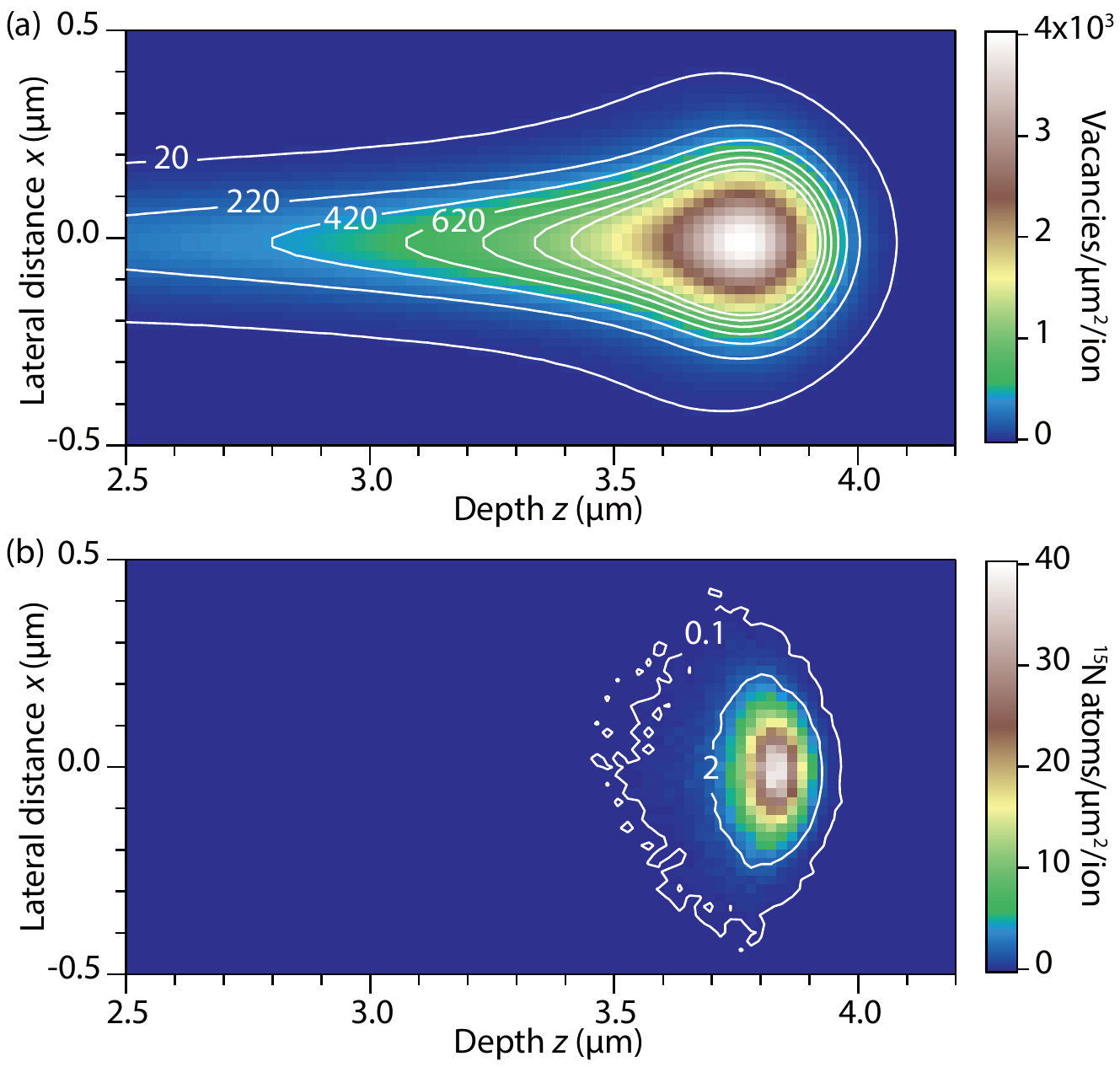}%
\caption{(Color online) Simulated $xz$-plane distributions of (a) vacancies with an isotropic diffusion length of $\sim 0.08\,\mu$m and (b) implanted $^{15}$N atoms with an incident energy of 10\,MeV, where $z$-axis is the implantation direction. The area-density isolines of vacancies and $^{15}$N atoms are shown as white lines in (a) and (b), respectively.~\label{FIG3}}
\end{figure}

Next we consider $^{14}$NV$^-$ formation. Preexisting $^{14}$N  impurities near an implanted $^{15}$N can be converted to $^{14}$NV$^-$ centers, as observed from the pairs of $^{14}$NV$^-$ and $^{15}$NV$^-$. In addition to this, shallower $^{14}$N impurities near vacancy cascades generated by $^{15}$N ion implantation, can be also transformed into $^{14}$NV centers, and the creation probability depends on concentrations of substitional $^{14}$N impurities ($[\textrm{N}_s^0]$) and vacancies ($c_v^{\prime}$). Hence, Eq.\,\ref{F1} can be modified as
\begin{equation}
Y([\textrm{N}_s^0], c_v^{\prime})=\frac{[\textrm{N}_s^0]\times V_v^{\prime}\times P_r}{N_i}\times P_t(c_v)\times\frac{c_v^{\prime}}{c_v},
~\label{F2}
\end{equation}
where $V_v^{\prime}$ is an effective volume containing enough vacancies to form $^{14}$NV$^-$ centers, and the replacement probability into carbon sites is given as $P_r=1$ since $^{14}$N impurities are substitutional atoms. The probability of trapping a vacancy, $P_t(c_v)\approx 1$ in Eq.\,\ref{F1}, is replaced by $P_t(c_v)\times \frac{c_v^{\prime}}{c_v}$, where the factor of $\frac{c_v^{\prime}}{c_v}$ results from a smaller vacancy concentration, $c_v^{\prime}$, for $^{14}$N impurities than $c_v$ for implanted $^{15}$N atoms. Figure\,\ref{FIG3}(a) and (b) show the simulated statistical distributions of vacancies (after diffusion) and $^{15}$N atoms, respectively, where $z$-axis is the ion implantation direction and the white solid lines indicate isolines of area density in each $xz$-plane projection. Calculating a volume of revolution ($V_v^{\prime}$) by rotating the area of more than 420\,vacancies/$\mu$m$^2$/ion around $x=0$ axis in Fig.\,\ref{FIG3}(a), and counting the number of vacancies ($n_v^{\prime}$) inside this volume, yields an average concentration of $c_v^{\prime}=\frac{n_v^{\prime}}{V_v^{\prime}}=28$\,ppb, where the selected depth range corresponds to the experimental range of $2.8\le\textit{z}\le3.9$ for $^{14}$NV$^-$ centers (Fig.\,\ref{FIG2}). Similarly, calculating the volume $V_v$ containing more than 2\,atoms/$\mu$m$^2$/ion from Fig.\,\ref{FIG3}(b), and counting the number of vacancies inside $V_v$ from Fig.\,\ref{FIG3}(a), gives the average concentration of $c_v=\frac{n_v}{V_v}=42$\,ppb. Here $V_v$ is the volume in which implanted $^{15}$N atom is found with the probability of 90\%. Assigning those values and the observed yield of $Y=0.56$ into Eq.\,\ref{F2}, we obtain a concentration of $^{14}$N impurities as 0.06\,ppb, which agrees with the expected value ($<0.1$\,ppb) from the growth condition.~\cite{Edmonds,Balasubramanian} 

Photoconversion between NV$^-$ and NV$^0$, observed in PL spectra has been reported for shallow NV$^-$ centers ($<200$\,nm depths), in spite of a similar ($\sim$ppb) or higher ($\sim$ppm) concentrations of N donors as described here.~\cite{Santori,Gaebel2,Ofori-Okai} Ionization of shallow NV$^-$ to the NV$^0$ charge state has been attributed to surface effects such as electron depletion due to an acceptor layer~\cite{Santori,Fu} or hole accumulation due to upward band bending at the hydrogen-terminated surface.~\cite{Hauf} The observation of the stable negatively-charged-state in the present study implies that surface effects are negligible for deep NV centers at 3-4\,$\mu$m depths. On the other hand, low energy implantation ($<5$\,keV)~\cite{Ofori-Okai} through nano-hole apertures~\cite{Toyli,Meijer2} is a promising route to fabrication of arrays of NV centers with high positional precision, however short $T_2$ times due to surface spins are problematic for building quantum registers. One solution is to overgrow an additional diamond layer onto the surface, which has succeeded in prolonging $T_2$ times.~\cite{Staudacher} Our results show that NV$^-$ spin qubits at the depth $\ge2.8\,\mu$m exhibit reliable properties of long coherence times and stable charge states.

In summary, engineered NV$^-$ spins qubits by $^{15}$N$^{3+}$ ion implantation into high-purity, $^{13}$C-depleted diamond were studied in a reduced background concentration of residual defects after high-temperature anneal. We observed a creation yield of $^{15}$NV$^-$ ($40\pm3\%$) which is likely to be limited by the population of implants having an interstitial configuration ($\sim60\%$).~\cite{Felton} Even with a N impurity concentration estimated as 0.06\,ppb, a considerable fraction of created NV$^-$ centers consisted of preexisting N impurities. The low concentration of nitrogen impurities, which allow for long coherence times ($\sim2$\,ms), still play a significant role for NV$^-$ charge stabilization. A provisional mechanism for charge stabilization will be required when fabricating NV$^-$ from only implanted ions by further lowering the N donor concentration.

This study was carried out as {\textquoteleft}Strategic Japanese-German Joint Research Project' supported by JST and DFG (FOR1482, FOR1493 and SFB716), ERC, DARPA and the Alexander von Humboldt Foundation. We thank Denis Antonov, Yoshiyuki Miyamoto, Brett C. Johnson, and Alexander P. Nizovtsev for valuable discussions, and Kay D. Jahnke, Pascal Heller, Alexander Gerstmayr, and Andreas H\"au{\ss}ler for assistance with the experiments.

\end{document}